\documentclass[12pt]{iopart}

\usepackage{cite}
\usepackage{hyperref}
\usepackage{braket}
\usepackage[utf8]{inputenc}
\expandafter\let\csname equation*\endcsname\relax
\expandafter\let\csname endequation*\endcsname\relax
\usepackage{amsmath}\newcommand{\bk}{\mathbf{k}}
\usepackage{amssymb}
\usepackage{graphicx}
\newcommand{\lambdabar}{{\mkern0.75mu\mathchar '26\mkern -9.75mu\lambda}}
\usepackage{hhline}
\newcommand{\noop}[1]{}

\begin{document}

\title[Quantum vacuum processes in the light of relativistic plasma mirrors]{Quantum vacuum processes in the extremely intense light of relativistic plasma mirror sources}

\author{Antonin Sainte-Marie}
\address{%
Lasers, Interaction and Dynamics Laboratory, Université Paris-Saclay, CEA, CNRS, LIDYL, 91191 Gif-sur-Yvette, France}%

\author{Luca Fedeli}
\address{%
Lasers, Interaction and Dynamics Laboratory, Université Paris-Saclay, CEA, CNRS, LIDYL, 91191 Gif-sur-Yvette, France}%

\author{Neïl Zaïm}
\address{%
Lasers, Interaction and Dynamics Laboratory, Université Paris-Saclay, CEA, CNRS, LIDYL, 91191 Gif-sur-Yvette, France}%

\author{Felix Karbstein}
\address{%

Helmholtz Institut Jena, GSI and
TPI, Friedrich-Schiller-Universit\"at Jena
}%

\author{Henri Vincenti}
\address{%
Lasers, Interaction and Dynamics Laboratory, Université Paris-Saclay, CEA, CNRS, LIDYL, 91191 Gif-sur-Yvette, France}%

\begin{abstract}
    The advent of Petawatt-class laser systems allows generating electromagnetic fields of unprecedented strength in a controlled environment, driving increasingly more efforts to probe yet unobserved processes through their interaction with the quantum vacuum. Still, the lowest intensity scale governing these effects lies orders of magnitude beyond foreseen capabilities, so that such endeavor is expected to remain extremely challenging. In recent years, however, plasma mirrors have emerged as a promising bridge across this gap, by enabling the conversion of intense infrared laser pulses into coherently focused Doppler harmonic beams lying in the X-UV range. In this work, we present quantitative predictions on the quantum vacuum signatures produced when such beams are focused to intensities between $10^{24}$ and $10^{28}$ $\textrm{W.cm}^{-2}$. These signatures, which notably include photon-photon scattering and electron-positron pair creation, are obtained using state-of-the-art massively parallel numerical tools. In view of identifying experimentally favorable configurations, we also consider the coupling of the focused harmonic beam with an auxiliary optical beam, and provide comparison with other established schemes. Our results show that a single coherently focused harmonic beam can produce as much scattered photons as two infrared pulses in head-on collision, and confirm that the coupling of the harmonic beam to an auxiliary beam gives rise to significant levels of inelastic scattering, and hence holds the potential to strongly improve the attainable signal to noise ratios in experiments.
\end{abstract}

\maketitle

\section{Introduction}
\label{Intro}

The macroscopic behavior of the electromagnetic field is remarkably well described by Maxwell's theory, which allows conceiving a vacuum state in which all properties of the field derive from laws devoid of reference to any other physical system~\cite{Landau1975_CFT}. Quantum field theory however reveals that all fundamental degrees of freedom come in the form of fields permeating all space and whose rest state can never be strictly assigned, so that an inert vacuum can in principle not exist nor a single field be in perfect isolation, each field constantly interacting with all the other ones through real or virtual states~\cite{Schwartz:2014sze}. As laser technology is making steady progress towards unprecedented intensities~\cite{PetaExaWW2019}, the interaction of light with the quantum vacuum newly stands as a promising probe for fundamental physics~\cite{Marklund2006_QEDrev,Dunne2008_ELI,HeinzlIlderton2009_QED_ELI_rev,Marklund2009_sfqed_las,Battesti2012_nl_vacQED,DiPiazza2012_rev,KingHeinzl2016_vacQED_las_rev}. According to the Standard Model, first deviations from free propagation stem from the coupling to the electron field, controlled by the electron mass $m$ and the coupling constant $e$. Together they define the characteristic field scale of Quantum Electrodynamics (QED), $F_S = m^2/e$ in Lorentz-Heaviside units with $c=\hbar=1$, that is $ F_S \simeq 1.32\times 10^{18} \text{ V\,m}^{-1} \simeq 4.4 \times 10^9 \text{ T}$, which marks the onset of the spontaneous coherent field decay into real electron-positron pairs known as the nonperturbative Schwinger process \cite{HE36,Schwinger51,Franklin91_QED_unstable_vacuum}. Much below this threshold though, dynamics still bears the mark of this coupling through virtual electron-positron pairs, giving rise to an effective electromagnetic self-interaction with a perturbative component only suppressed by powers of the field strength $F/F_S$ \cite{HE36,RITUS1972_rad_corr,Bialynicki-Birula1970,Dittrich1985_eff_L_qed,Dittrich2000_probe_QED}. Since the strongest macroscopic fields achievable to date \cite{Yoon21}, provided by femtosecond multi Petawatt-class lasers focused near diffraction-limit, reach $F_\mathrm{las} \sim 10^{-3} \times F_S$, direct observation of fully nonperturbative vacuum processes such as pair creation seems precluded in a foreseeable future, while research efforts devoted to light-by-light scattering have pointed out the yet elusive nature of this process together with the exciting possibility to attain a discernible signal in several optimized configurations \cite{LittGF_1,LittGF_2,LittGF_3,LittGF_4,LittGF_5,LittGF_5.5,LittGF_6,LittGF_7,LittGF_8,LittGF_9,LittGF_10,LittGF_11,LittGF_12,LittGF_13,LittGF_14,LittGF_15,LittGF_16,LittGF_18,LittGF_19,sangal2021} (see \cite{karbstein2020_particles} for a recent introduction and references therein). These field configurations typically involve multiple colliding intense optical pulses, or in some cases coupling to a high-energy electron or photon beam, providing existing and upcoming facilities with challenging experimental programs. 

In recent years, however, alternative paths have been breached to much stronger macroscopic fields \cite{PetaExaWW2019,gonoskov2021charged}. 
Among them, the so-called "Coherent Harmonic Focusing" (CHF) suggests to leverage on the Doppler frequency upshift of a high-power infrared laser upon reflection off a curved relativistic plasma mirror (PM), to compress its energy down to considerably smaller spatio-temporal volumes than the ones allowed with the larger original laser wavelength \cite{Bulanov1994,Bulanov2003_chf_scw,Gordienko2004_pm_prl,Naumova2004,BaevaGordienko2006_pm_theory,Solodov2006,Dromey2009,Teubner2009,AnDerBraegge_2010,Vincenti2014_optics_pm,Gonoskov2018}.

Extensive theoretical and numerical studies recently proposed realistic paths to implement such curved relativistic PMs, with several techniques regarding the focusing of the harmonic beam~\cite{Gordienko2005,Gonoskov2011_pm_ehi,Vincenti2019_prl,quere_vincenti_2021,Vincenti_PMopt}. These studies robustly predict focused intensities ranging from $10^{24}$ W\,cm$^{-2}$ to more than $10^{28}$ W\,cm$^{-2}$ with multi-Petawatt class laser systems. One of these techniques, which leverages the PM curvature induced by the radiation pressure of the incident laser \cite{Vincenti2019_prl}, has just found promising early experimental validation in the 100 TW regime \cite{Chopineau2021_st_charac_pm_pls}.  

Despite these remarkable figures and a context of rising interest in high-field physics at the Schwinger limit (see \emph{e.g.} \cite{BucksbaumLOI2020}), these recently proposed PM-based configurations have been comparatively little studied as sources for subsequent interactions, most probably due to their considerable numerical cost -- as an inherently many-scale problem -- and theoretical complexity -- as the product of highly non-linear laser-plasma interactions. 

In this work, we present quantitative predictions on the quantum vacuum signatures produced by these extremely intense light sources. To achieve this, we developed and benchmarked at very large scale state-of-the-art numerical tools able to realistically simulate such field configurations \cite{Asm_numerics}. 

In the following, we first detail the computational methods in sec.~\ref{secII}, consisting for the most part in a parallel numerical implementation of the Stimulated Vacuum Emission theoretical framework introduced by the authors of \cite{FK_SVE2015}, which we briefly recall.

The results obtained with our numerical code applied to PM-generated Doppler harmonic beams are then presented in sec.~\ref{secIII}. In order to gain theoretical insights into the photon signal dependence upon the harmonic beam spectrum, the PM beams results are compared with idealized harmonic beams made of $1\leq n_h \leq 16$ harmonics of a fundamental frequency of equal amplitude at focus, and supplemented with analytical predictions derived from a simple model of these field configurations.

Beyond maximizing the number of scattering events, in practice all-optical experiments for probing the quantum vacuum have to confront with the problem of clearly assessing the quantum origin of the detected photons. This may prove all the more crucial in the CHF configuration due to the possibly abundant emission of ”background photons" of non-quantum origin during the laser-plasma interaction. In view of identifying ways to relax the associated experimental constraints, two hybrid scenarios are finally presented in sec.~\ref{secIV}, where the Doppler harmonics field plays the role of an ultra-intense pump while the polarized vacuum is probed by a controlled auxiliary source. We considered for the probe either a mildly focused PW-class infrared pulse, or a 100 TW-class green laser beam.

The conclusions of this work are exposed in sec.\ref{Concl} along with a brief outlook.

\section{Vacuum photon emission algorithm}
\label{secII}

\subsection{Formalism}

A classical field configuration is naturally embedded in Quantum Electrodynamics under the form of a coherent state, and in free theory this correspondence commutes with time evolution \cite{Glauber_ch2,Reuse}. In particular, expectation values of the photon numbers in each mode, of momentum $\mathbf{k}$ and polarisation $p$, are encoded \emph{via} the modulus of the Fourier coefficients of the classical field, which directly implies they are conserved quantities under Maxwell's action. Together with its experimental readiness, this property makes photon numbers an observable suited for detecting couplings to the quantum vacuum, and we therefore wish to compute
\begin{equation}
    d^3N_{(p)}(\bk) = \braket{ n_{(p)}(\mathbf{k})}_{\chi_\mathcal{A}} - d^3N^\mathcal{A}_{(p)}(\bk)
\label{Sve}
\end{equation}
the change in the expectation value of the photon number operator $n_{(p)}(\mathbf{k})$ measured at asymptotically late times taking the coherent state $\chi_\mathcal{A}$ as the initial state, compared to the initial photon content $d^3N^\mathcal{A}_{(p)}(\bk)$ associated to the classical field with 4-potential $\mathcal{A}$. Following \cite{FK_SVE2015}, we discard the marginal effects of coherent multi-photon emission and interference with the classical field to obtain
\begin{align}
    d^3N_{(p)}(\bk) &\equiv \frac{d^3k}{(2\pi)^3}\left\vert S_{(p)}(\bk) \right\vert^2\\
    \text{with, } S_{(p)}(\bk) &= \braket{\gamma_{(p)}(\bk) |\Gamma_\mathrm{int}^\mathcal{A}\left[a(x) \right] |0 }
\end{align}
where $\ket{\gamma_{(p)}(\bk)} = a^\dagger_{(p)}(\bk)\ket{0}$ is the $(\bk,p)$ mode one-photon state vector, and $\Gamma_\mathrm{int}^\mathcal{A}\left[a(x) \right]$ the effective interaction operator encoding couplings of the dynamical (operator-valued) photon field $a(x)$ with other fields for the initial state $\chi_\mathcal{A}$. 

From orthogonality of the photon basis we can furthermore write \cite{FK_AllOpt2018},
\begin{equation}
    S_{(p)}(\bk) = \Braket{\gamma_{(p)}(\bk) | \left.\int d^4x \frac{\delta \Gamma^\mathcal{A}_\mathrm{int}}{\delta a^\mu(x)}\right\vert_{a=0} a^\mu(x) | 0 }
\end{equation}
or expressing $a^\mu$ in Lorenz gauge,
\begin{equation}
    S_{(p)}(\bk) = \frac{\varepsilon^{\mu \ast}_{(p)}(\bk)}{(2\pi)^3\sqrt{2\mathrm{k}}} \int d^4x  e^{-i(\mathrm{k} t-\bk\cdot\mathbf{x})} \left. \frac{\delta \Gamma^\mathcal{A}_\mathrm{int}}{\delta a^\mu(x)}\right\vert_{a=0} \text{, for } \mathrm{k} = \lVert \mathbf{k} \rVert
\end{equation}
Within QED, in the limit of slowly varying classical electromagnetic fields, retaining only the dominant interaction with the electron field then leads to the identification of $\Gamma_\mathrm{int}^\mathcal{A}$ with the 1-loop Heisenberg-Euler action \cite{HE36,Schwinger51,FK_Addendum}, at first order in $\alpha = e^2/4\pi$ and nonperturbatively in the classical field strength $F_\mathcal{A}$. In such fields varying on characteristic spatio-temporal scales $$\lambda\gg1/m$$  subleading derivative corrections which scale as $\sim1/(m\lambda)^2$ can be safely neglected, such that $$\Gamma_{\rm int}^{\cal A}\simeq\int d^4x {\cal L}_{\rm int}^\mathrm{HE}(F)|_{ F\to  F_\mathcal{A}(x)+f(x)}$$ with constant-field Heisenberg-Euler interaction Lagrangian ${\cal L}_{\rm int}^\mathrm{HE}$ and quantized photon field $f$. Taking as polarization vectors $\varepsilon^\mu_p(\bk) = (0,\mathbf{e}_p(\bk))$ with $p\in\lbrace 1,2\rbrace $ such that $(\bk/\mathrm{k},\mathbf{e}_1(\bk),\mathbf{e}_2(\bk))$ forms a direct orthonormal basis, the amplitudes reduce to
\begin{equation}
    S_{(p)}(\bk) = i\sqrt{\frac{\mathrm{k}}{2}}\int d^4x e^{-i(\mathrm{k}t-\bk\cdot\mathbf{x})}\left[ \mathbf{e}_p(\bk)\cdot \mathbf{P}(x) + \mathbf{e}_{p+1}(\bk)\cdot \mathbf{M}(x) \right] 
\label{Tf_PM}
\end{equation}
with formal convention $\mathbf{e}_{p+2} = -\mathbf{e}_p$, and defining $\mathbf{P} = \frac{\partial \mathcal{L}_\mathrm{int}^\mathrm{HE}}{\partial \mathbf{E}}$ and $\mathbf{M} = \frac{\partial \mathcal{L}_\mathrm{int}^\mathrm{HE}}{\partial \mathbf{B}}$ the induced vacuum polarisation and magnetisation.

We finally restrict ourselves to field strengths $F \ll F_S$, so that up to $\mathcal{O}(F^2/F_S^2)$ corrections we have,
\begin{equation}
    \mathcal{L}_\mathrm{int}^\mathrm{HE} \simeq \frac{m^4}{8\pi^2}\left(\frac{e}{m^2} \right)^4 \left[ 4\mathcal{F}^2 + 7\mathcal{G}^2 \right]
\label{Lag}
\end{equation}
where $ \mathcal{F} = \frac{1}{2}(\mathbf{B}^2-\mathbf{E}^2) \text{ and } \mathcal{G} = -\mathbf{B}\cdot\mathbf{E}$ are the gauge- and Lorentz-invariants of the field.

The number of Schwinger pair creation events was besides estimated assuming no particles-field retro-action, by integrating the local production rate derived from the nonperturbative form of $\mathcal{L}^ \mathrm{HE}_\mathrm{int}$ in the simulated space-time volume, as proposed \emph{e.g.} in ref. \cite{Bulanov2006_EPpairs,Gavrilov2017}:
\begin{equation}
    N_S \equiv \frac{1}{4\pi^2} \lambdabar_C^{-4} \int d^4x\ \epsilon\eta \coth\left(\frac{\pi\eta}{\epsilon}\right)\exp\left(-\frac{\pi}{\epsilon}\right)
    \label{Scw_rate}
\end{equation}
where $\lambdabar_C = \hbar/mc$ is the reduced Compton length of the electron and
\begin{align}
    & \epsilon = \frac{1}{F_S} \sqrt{\sqrt{\mathcal{F}^2+\mathcal{G}^2}+\mathcal{F}} \text{, } \eta = \frac{1}{F_S}\sqrt{\sqrt{\mathcal{F}^2+\mathcal{G}^2}-\mathcal{F}}
\end{align}
The conditions of validity of this approach are further discussed in sec.~\ref{secIII}.

\subsection{Numerical implementation}

The accurate estimation of the quantities of interest Eqs.~\ref{Sve} and \ref{Scw_rate} for arbitrary field configurations can be performed numerically. In order to achieve this we developed and optimized a code able to run on large scale parallel computing infrastructures, whose detailed presentation and validation can be found in \cite{Asm_numerics}. 

Here, we give the general principle of this algorithm, similar in nature to the one presented in \cite{FK_AllOpt2018,FK_AllOpt2019}. Let us focus on the computation of Eq. \ref{Tf_PM}, which amounts to a four dimensional on-shell ($k^0=\lVert\mathbf{k}\rVert$) Fourier transform of third degree polynomials in the $\mathbf{E}$ and $\mathbf{B}$ fields values. We numerically estimate it by performing 3D space parallel fast Fourier transforms of the sources $\mathbf{P}$ and $\mathbf{M}$ from position to momentum space at each timestep $t_j$, followed by a 1D time integration of all these contributions multiplied by the appropriate phase factor, namely
\begin{align}
    S_{(p)}(\bk) &\equiv i\sqrt{\frac{\mathrm{k}}{2}} \left[ \mathbf{e}_p(\bk)\cdot \widetilde{\mathbf{P}}(\bk) + \mathbf{e}_{p+1}(\bk)\cdot \widetilde{\mathbf{M}}(\bk) \right]  \\
    \text{where \emph{e.g.} } \widetilde{\mathbf{P}}(\mathbf{k}) &\simeq \sum_{j=0}^{N_t} e^{-i\mathrm{k}t_j} \mathrm{FFT_3}\left[\mathbf{P}(\mathbf{E}(t_j),\mathbf{B}(t_j))\right] \Delta t
\end{align}

Field values at time $t_j$ are accessed by exact Maxwell propagation of an initial field configuration performed \emph{via} 3D spatial Fourier transforms (plane wave decomposition). The computational cost of these simulations stems from the inherently tridimensional nature of the source fields $\mathbf{P}$ and $\mathbf{M}$, and from the necessity to resolve all spacetime scales from the total beam focusing length and pulse duration to the highest frequency content of the field. In the case of the harmonic beams, this imposes the use of a supercomputer (see \ref{App_comp} for details on the computational resources).

The above described procedure involves an initialisation step, consisting in defining the field configuration in space at some given time and computing its amplitude for a given total energy. The careful definition of this initial condition is all the more important because, as explicit in Eq.~\ref{Lag}, QED effects depend on the Lorentz invariants of the field, which can strongly reflect even minor variations in the field structure, as in particular they would strictly vanish for a plane wave. The investigation of the robustness of the simulation results upon the numerical procedure details and involved approximations was carried out in \cite{Asm_numerics}. In this work, initial beam configurations were defined by $A(\omega)$, the time spectrum of their main field component on-axis in a given plane $x=x_0$, under the assumption that each monochromatic component has a Gaussian spatial profile, given by the paraxial expressions of the Gaussian beam $E_{px}(\mathbf{x},\omega)$ \cite{Salamin2006_fields_of_a_gaus} and characterized by its waist and radius of curvature in plane $x=x_0$. The initial field is then reconstructed by summing over all monochromatic contributions, $E(\mathbf{x},t_0)\equiv \sum_{\omega} e^{-i\omega t_0} E_{px}(\mathbf{x},\omega)A(\omega)$. The above field profiles being built from paraxial expressions, they do not exactly satisfy vacuum Maxwell's equations, we hence made the choice to project them on the exact solution space by dropping all longitudinal field components (\emph{i.e.} making it divergence free) \emph{via},
\begin{align}
    \left\lbrace
    \begin{aligned}
        \mathbf{E}(\bk) &\equiv a_1(\bk) \mathbf{e}_1(\bk) + a_2(\bk) \mathbf{e}_2(\bk)\\
        \mathbf{B}(\bk) &\equiv a_1(\bk) \mathbf{e}_2(\bk) - a_2(\bk) \mathbf{e}_1(\bk)
    \end{aligned}
    \right.
    \text{ , with }
    a_p(\bk) = \mathbf{e}_p(\bk)\cdot \mathbf{E}(\bk) 
\end{align}

Several observables of interest are finally extracted from the photon emission amplitudes that will be used as figures of merit for the different configurations: 
\begin{itemize}
    \item $N_t$, the total number of scattered photons 
    \item $N_\perp$, the total number of photons emitted with polarization crossed to the main polarization direction of the driving beam, typically identifiable as its polarization direction at focus in all cases we consider
    \item $N_{t,\perp}^{>}$, the corresponding quantities of \emph{discernible} photons, generically defined as those emitted outside of the opening cones of the driving beams, as commonly done in this context \cite{FK_SVE2015,FK_AllOpt2018,FK_Multicolor}, with some more detailed criteria that will be specified in each case of study.
\end{itemize}

\section{Vacuum signatures from a coherently focused harmonics field}
\label{secIII}

\begin{figure}[t!]
\begin{center}
\includegraphics[width=1.0\columnwidth]{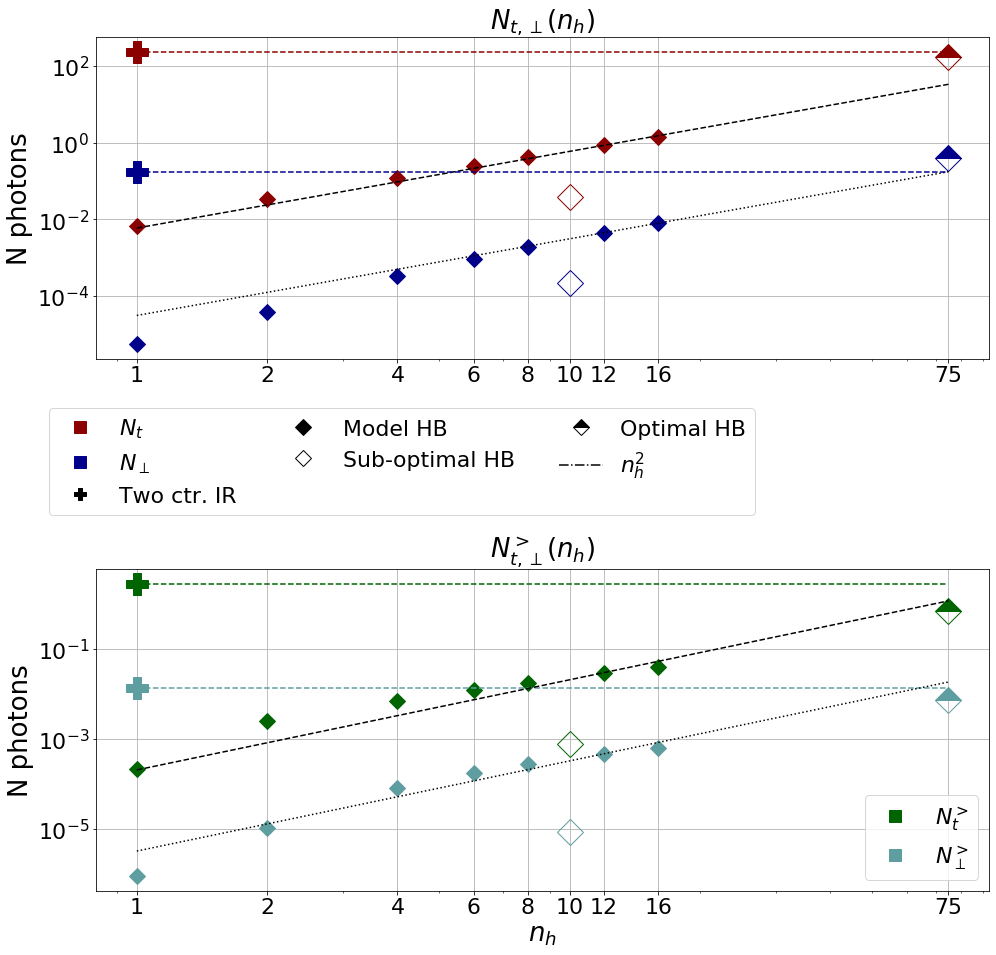}
     \vspace{-20pt}
     \caption{\textbf{Integrated photon numbers as a function of $n_h$ for a model beam made of $n_h$ harmonics of its fundamental frequency, with quasi-equal diffraction parameter $\varepsilon$, focal point and amplitude at focus, or PM-generated beams of effective number of harmonics $n_h$.} The case of two counterpropagating fundamental frequency pulses \textbf{(crosses)} is shown at $n_h=1$ for comparison. \textbf{(top)} Total number of photons $N_t$ or with crossed-polarization $N_\perp$. \textbf{(bottom)} Total numbers of discernible photons, emitted outside of the $\theta = \arcsin(2\varepsilon)$ cone directed along the beam propagation direction. All beams are at diffraction limit ($\varepsilon=1/\pi$).
     }
     \label{Fig3} 
\end{center}
\end{figure}

We now turn to the study of PM-generated fields as sources for the quantum vacuum processes of photon-photon scattering and Schwinger electron-positron pair creation. In the interaction of a relativistic-intensity laser (normalized vector potential $a_0 = eE_0/(mc\omega_0)>1$) with a sharp solid-density plasma surface, the electron population effectively acts as a "relativistic oscillating mirror" and thus converts by Doppler effect the incident laser pulse into a train of sub-wavelength pulses, equivalently described as a coherent superposition of harmonics of the initial laser beam, with effective harmonics orders up to $n_h\lesssim100$ \cite{Bulanov1994,BaevaGordienko2006_pm_theory,Teubner2009,Vincenti2014_optics_pm,Gonoskov2018,Vincenti2019_prl}. Although crucial to many applications including the present one, the analytical modelling of such fields remains very challenging to date, especially regarding the correct prediction of the spectrum amplitude and phase, and therefore of the associated maximum attainable field amplitude. We addressed this difficulty via first-principles kinetic Particle-In-Cell (PIC) simulations \cite{Arber_2015} of the harmonics time spectrum at a point near its region of generation. The numerical data obtained from these simulations allows us to fully reconstruct a realistic 3-dimensional harmonic field, under the hypotheses that \emph{i)} the spatial profiles $E(\mathbf{x},\omega)$  of the harmonic beams are well-described by the Gaussian paraxial field expressions at each frequency $\omega$, and \emph{ii)} all frequencies share the same waist $w_p$ and radius of curvature $R_p$ in the plane of generation, set to coincide with the plasma mirror plane of the original PIC simulation (see \emph{e.g.} \cite{Vincenti2019_prl,Vincenti_PMopt} for a motivation from simulation results). 

We studied two distinct harmonic fields, corresponding to two different conditions of generation of their original time spectrum. They are both generated by a multi-PW class laser of normalized potential $a_0\sim 75$ on the PM surface, with an optimal vacuum-plasma density gradient (see \cite{Vincenti2019_prl,Edwards2020,quere_vincenti_2021,Vincenti_PMopt} for details on optimal harmonics generation), and only the incidence angle is varied. The choices employed here are either 45° or 55°. As the latter angle is found to be an optimum, we refer to the associated fields as optimal harmonic beams. On the other hand, an angle of 45° gives rise to a sub-optimal harmonic beam (see \ref{App_harmonics}). In all studied cases, the initial field configuration was defined in the PM plane using zeroth order paraxial expressions \cite{Salamin2006_fields_of_a_gaus} before projection (\emph{cf.} sec.~\ref{secII}), and the total energy in the harmonic beam was set to $W=50$ J. 

\subsection{Photon scattering}

For a single beam of frequency $\mathrm{k}_0$, it has been shown \cite{FK_AllOpt2019,MondenKodama2011} that the number of vacuum scattered photons increases with tighter focusing, demonstrating that the induced increase in field invariants overcomes the decrease in interaction volume. More specifically, analytical calculation of the photons emitted by a single frequency focused Gaussian pulse gives:
\begin{equation}
    N_{t,\perp} \simeq C_{t,\perp} \left(\frac{\mathrm{k}_0}{m} \right)^3\frac{1}{(m\tau)^2}\left(\frac{W}{m} \right)^3\varepsilon^8
    \label{N_al_gaus}
\end{equation}
where $C_{t,\perp}$ is a polarization-dependent constant, $m$ the electron mass, $\tau$ its duration assuming a time profile of the form $E(t)=E_0 e^{-\frac{(t-x)^2}{(\tau/2)^2}}$, $W$ its total energy, and $\varepsilon = 2/\mathrm{k}_0w_0$ its diffraction parameter. The diffraction parameter being physically constrained to $\varepsilon\lesssim 1/\pi$, converting the beam to higher frequencies then appears as a straightforward way to increase the scattered photon number. The technique considered in this study for physically achieving this conversion involves generation of high-harmonics of the initial frequency, so that the energy ends up distributed among many different frequencies. We can refine the calculation to account for this multiple frequency content along the line of \cite{BoostCHF2019}. This is done by assuming that $E_h = \sum_{n=1}^{n_h} E_{0n}$ with $E_{0n} = \sqrt{\frac{\delta W_n}{\tau w_{0n}^2}} = E_0^g\frac{n^{-s+1}}{(H_{2s}(n_h))^{1/2}}$ for a fundamental frequency Gaussian pulse amplitude $E_0^g=\sqrt{\frac{ W}{\tau w_{01}^2}} $ and $H_q(n) = \sum_{m=1}^n m^{-q}$, where we have introduced the total beam duration $\tau$, energy $W$, roll-off parameter of the power law spectrum $s$, and energy fraction $\delta W_n$ and waist at focus $w_{0n}$ of harmonic component of order $n$. Besides, in the limit $n_h\gg1$ and $s \lesssim 2$ the field profile takes the form of a train of sub-wavelength pulses, whose individual profiles can then be approximated as Gaussian with $\tau_h \sim 1/n_h\mathrm{k}_0$, $w_{h} \sim w_0/n_h$, leading to
\begin{equation}
    N_{t,\perp}^{h} = N_{t,\perp}^g\frac{1}{\left(H_{2s}(n_h)\right)^3}\left(\frac{H_{s-1}(n_h)}{n_h}\right)^4\times \left(H_{s-1}(n_h)\right)^2
\end{equation}
with $N_{t,\perp}^g$ the photon number corresponding to a fundamental frequency Gaussian beam of same duration and energy up to a constant proportionality factor. In the case of optimal PM high-harmonics generation we have $s\simeq 1$, so that,
\begin{equation}
    N_{t,\perp}^h \propto n_h^2.N_{t,\perp}^g
    \label{CHF_nh_scale}
\end{equation}
where $n_h \sim 100$, implying this method can \emph{a priori} enhance the emitted photons number by 4 orders of magnitude over a single tightly focused infrared laser beam of same duration and energy.

As a first test of the viability of this prediction for PM-generated harmonics, we consider the simpler case of $n_h$ harmonics at $s=1$, \emph{i.e.} such that the amplitude of each harmonic is equal at focus (\emph{cf.} for instance \cite{Vincenti2019_prl,Vincenti_PMopt} and \ref{App_harmonics} for comparison with PM-generated spectra) for $n_h$ ranging from 2 to 16. In view of the CHF scenario we simply considered that all harmonics are perfectly in phase in some plane at a finite distance from focus, and share the same waist $w_p = 5\lambda_0$ and radius of curvature $R_p$ in this plane, so that they all approximately have the same diffraction parameter and focal point provided focusing is strong enough ($w_0(\lambda_0)^2/w_p^2 \ll 1$). The global duration and energy of the model beam are taken respectively as $\tau = 25$ fs (in practice by defining an $A(\omega)$ with peaks of appropriate spectral width, \emph{cf.} sec.~\ref{secII}), chosen close to the effective duration of the harmonics pulses train, and $W = 50$ J.

In Fig.~\ref{Fig3} we plot the number of scattered photons as a function of the harmonic number $n_h$ for fixed other parameters. In particular, we take $R_p = \pi w_p$ so that all frequencies $\omega$ satisfying $\omega w_p / 2 \gg 1$ are at diffraction limit. Besides, we define discernible photons as those scattered at an angle $\theta = \sphericalangle (\mathbf{e}_x,\bk)$ with the harmonic propagation direction greater than $\theta_2 = \arcsin(2\varepsilon)$, which corresponds to a background laser photon density drop off of at least $1/e^8$ in a Gaussian beam model. As introduced in Sec.~\ref{secII}, the corresponding numbers are denoted by a superscript $>$. The results confirm the increase in signal photons with the increase of harmonics number, and the scaling law reported in Eq.~\ref{CHF_nh_scale} at least for the total photon numbers. The crossed-polarized photons display an even faster increase with $n_h$ at first, before converging to the $n_h^2$ trend. At $n_h=16$, photon numbers reach $N_t = 1.443 \ (N_\perp=8.152\times 10^{-3})$, and $N^{>}_t = 3.955\times10^{-2} \ (N^{>}_\perp=6.311\times10^{-4})$, demonstrating enhancement factors of $a_t \equiv N_t/N_t^{n_h=1} = 2.1\times10^2 \ (a_\perp \equiv N_\perp/N_\perp^{n_h=1} = 1.5\times10^3 )$ for the total photon numbers, and  $a^{>}_t = 1.8\times10^2 \ (a^{>}_\perp = 7.0\times10^2 )$ for the discernible signal. We note that the smaller increase of the discernible signal may be due to the discernibility criterion invoked here: as high-frequency photons tend to be more collimated, a frequency-independent exclusion cone generically disfavors configurations with a substantial high-frequency energy content. 

For comparison, we also show the photon numbers attainable with PM-generated harmonic beams created in sub-optimal or optimal laser-plasma interaction conditions. In both cases we define their associated effective harmonics number as $n_h \equiv \sum_{n=1}^{\infty} \vert E_{foc}(n\omega_0) \vert/\vert E_{foc}(\omega_0)\vert $ with $E_{foc}$ the amplitude of harmonic order $n$ at focus, consistently with the $s=1$ hypothesis. 
This already reveals that while sub-optimal CHF (sub-optimal HB in Fig.~\ref{Fig3}) allows for enhancement factors of $a_t = 5.9 \ (a_\perp = 3.9\times 10^1 )$ and  $a^{>}_t = 3.5 \ (a^{>}_\perp = 9.7 )$, it falls well below the values expected from our model due to the fast decrease of the spectrum in this case. On the opposite, the optimal harmonic beam results (optimal Hb in Fig.~\ref{Fig3}) are much more closely in line with Eq.~\ref{CHF_nh_scale}, and even exceed it in total photon numbers, as a consequence of the slight growth of the harmonics amplitudes at focus for the first dozen of harmonics orders (\emph{cf.} \ref{App_harmonics}). The corresponding enhancement factors are $a_t = 2.5\times 10^4 \ (a_\perp =7.3\times 10^4 )$ and  $a^{>}_t = 3.1\times 10^3 \ (a^{>}_\perp = 8.0\times 10^3 )$, demonstrating the efficiency of CHF under these conditions, and therefore the importance of optimizing harmonic generation to obtain good photon signal. The resulting total photon numbers for a diffraction limited focusing ($\varepsilon = 1/\pi$) are respectively $N_t = 4.738\times10^{-2} \ (N_\perp = 6.047\times10^{-3} )$ for the sub-optimal beam, and $N_t = 170.3 \ (N_\perp = 4.088 \times 10^{-1})$ for the optimal beam. The discernible photon numbers are $N_t^{>} = 7.580\times10^{-4} \ (N_\perp^{>} = 8.696\times10^{-6} )$ for the sub-optimal beam, and $N_t^{>} = 1.110 \ (N_\perp^{>} = 4.777\times10^{-2} )$ for the optimal beam, which implies about one discernible photon per shot.

\begin{figure}[t!]
\begin{center}
\includegraphics[width=1.0\columnwidth]{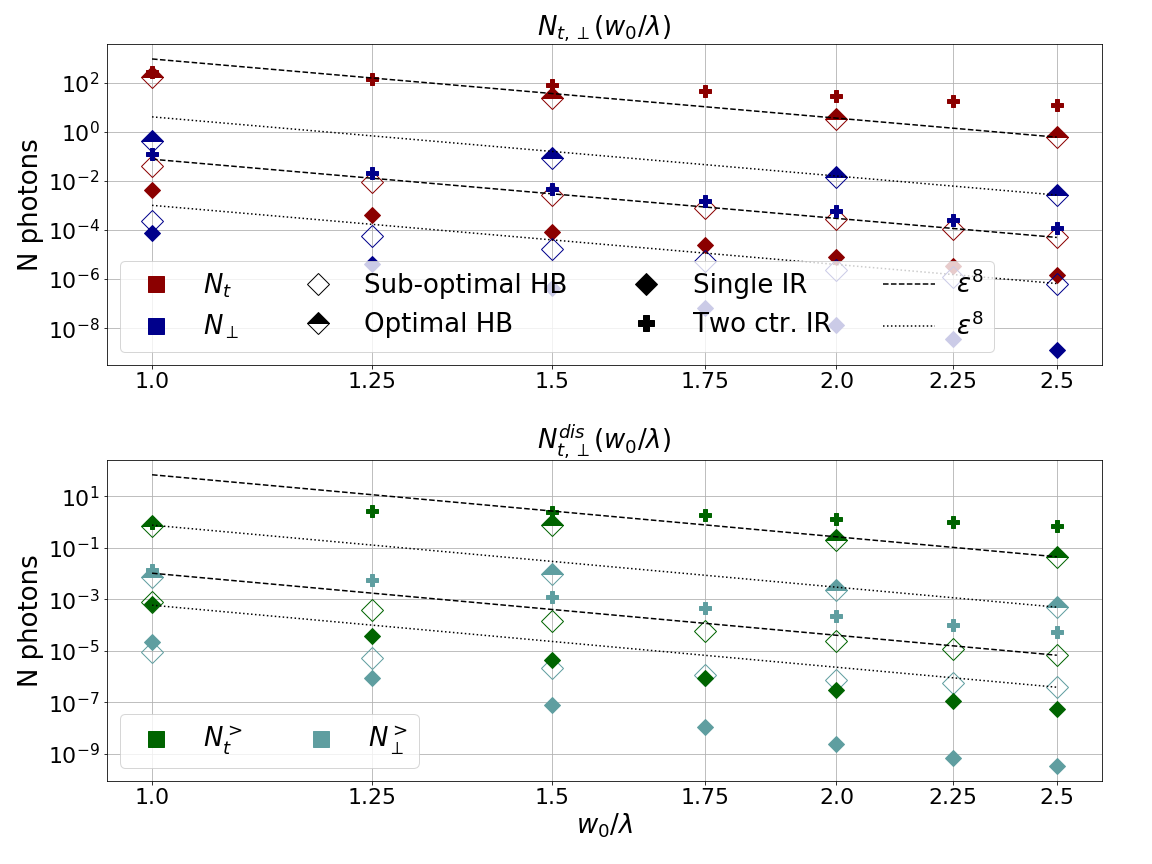}
     \caption{\textbf{Integrated signal photon numbers as a function of $w_0/\lambda$, the waist to wavelength ratio common to all frequencies. Here we present results for a relativistic PM-generated coherent harmonic beam (HB), a single fundamental frequency pulse (Single IR) and two counter-propagating fundamental frequency pulses (Two ctr. IR), focused in a vacuum.} \textbf{(top)} Total number of photons $N_t$ or with crossed-polarization $N_\perp$.  \textbf{(bottom)} Total numbers of discernible photons, emitted outside of the $\theta = \arcsin(2\varepsilon)$ cone directed along the beam propagation direction (both directions for \textbf{(Two ctr. IR)}).
     }
     \label{Fig6}
\end{center}
\end{figure}

Results regarding the photon numbers the dependence of the photon numbers on the beam  diffraction parameter $\varepsilon$ are plotted in Fig.~\ref{Fig6}. In all considered cases, the frequency-dependent waist at focus satisfies $w_0(\omega)\ll w_p$ for all frequencies, from which we can derive $\varepsilon = 1/\pi(w_0/\lambda) \simeq w_p/R_p$ so that the diffraction parameter is indeed nearly the same for all frequencies. We observe manifestly the same trends as in the cases of a single Gaussian beam~\cite{FK_AllOpt2019,MondenKodama2011}, namely a scaling with $\varepsilon^8$ of all photon numbers. This is fully consistent with the above described analytical estimates Eq.~\ref{N_al_gaus} for the single Gaussian beam, and Eq.~\ref{CHF_nh_scale} derived from a Gaussian pulse model of the individual harmonics pulses.

For reference, we can compare these results to the one obtained for the same total energy with two fundamental frequency pulses in head-on collision, each carrying half the total energy (Two ctr. IR in Fig.~\ref{Fig3}). The importance of this configuration becomes clear noticing that the field invariants entering the Heisenberg-Euler Lagrangian Eq.~\ref{Lag} identically vanish for a plane wave, making the field of a single loosely focused laser \emph{a priori} very sub-optimal for observing quantum vacuum effects. The transient standing wave obtained simply by crossing two such laser pulses on the opposite yields near maximal invariants for the same total electromagnetic energy, and therefore provides a landmark for the achievable signal in all-optical frequencies setups. With our chosen parameters (50 J total energy, 25 fs duration, $\varepsilon = 1/\pi)$), the resulting values of the total and discernible photon numbers are $N_t = 241.1\ (N_\perp = 1.176\times10^{-1})$ and $N^{>}_t = 2.756\ (N^{>}_\perp = 1.321\times10^{-2})$.

In the optimal case, it therefore appears that hat although a single beam photon signal is generically suppressed by at least $\varepsilon^4$ compared to counter-propagative beam configurations, individual PM-generated beams reach comparable signal levels, and hence allow bypassing the fine alignment and synchronisation conditions required to obtain sizable signals in multi-beam configurations. The final numbers remain admittedly low, so that the costs and benefits of each approach still has to be carefully determined. In particular, CHF alleviates the constraint of spatial and temporal synchronization at the micron and fs level of multiple focused high-power laser beams, demanding instead a fine control of the laser-plasma interaction conditions. 
While promising paths are being explored to this end, it may still happen that the background radiation inherent to the harmonics-generating plasma excitation overcomes the increase in signal photons, precluding a conclusive vacuum photon detection in the harmonic beam direction. In order to overcome this potential limitation, in sec.~\ref{secIV} we present a compromise scenario close to the one put forward in \cite{BoostCHF2019}, where a second beam is used to build a vacuum photon signal out of the plasma radiation direction, while we leverage the extreme harmonic beam intensity to relax the focusing and intensity constraints on this secondary beam.

\subsection{Schwinger pair creation}

\begin{figure}[h!]
\begin{center}
\includegraphics[width=1.0\columnwidth]{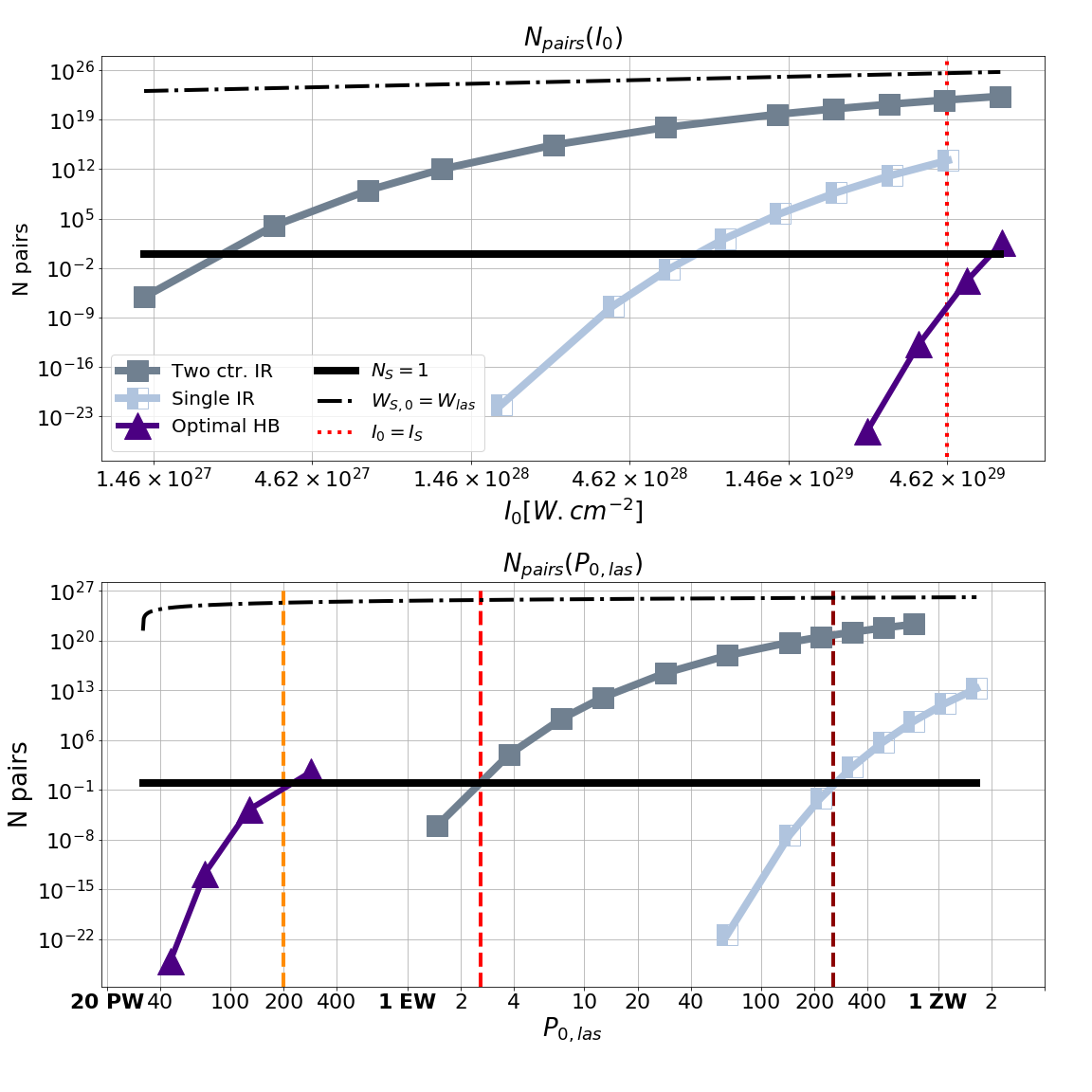}
    \vspace{-20pt}
     \caption{\textbf{Number of Schwinger electron-positron pairs created neglecting particles-field retroaction.} \textbf{(top)} Number of Schwinger pairs created as a function of peak intensity for the field configurations of two fundamental frequency pulses in head-on collision \textbf{(Two ctr. IR)}, a single one \textbf{(Single IR)} and an optimal PM-generated harmonic beam \textbf{(Optimal HB)}; all beams are focused at diffraction limit. The black lines show the thresholds of one created pair (solid) and pairs number corresponding to total field conversion to pairs mass energy (dashed). \textbf{(bottom)} Number of Schwinger pairs created as a function of driving beam power. In the harmonic beam scenario, power values stand for the laser incident on plasma mirror surface, accounting for reflection coefficients $R_{pm}$ set to the realistic value of $70\ \%$ \cite{Vincenti_PMopt}.
     }
     \label{Fig7}
\end{center}
\end{figure}

The peak intensity obtained when focusing the optimal harmonic beam with $100$ J total energy is close to $1.4\times10^{28} \mathrm{\ W\,cm^{-2}}$, corresponding to field amplitudes of $17\ \%$ of the Schwinger critical value. At such scales it becomes relevant to investigate electron-positron pair creation via the Schwinger process. Although strictly speaking the vacuum QED formalism used in this work does not hold in the event of one or more pair creation, requiring \emph{e.g.} a self-consistent semi-classical evolution of fields and particles as done by PIC-QED algorithms \cite{Gonoskov2015_PICreview,Zoni2021}, it nevertheless allows determining the number of pairs produced assuming no field-particles retro-action. Under the same hypotheses as for vacuum photon emission, together with the Keldish adiabatic condition $\gamma_K = 1/m l_E F_0^{-1} \ll 1$ \cite{Dunne2009,FK_HE2017}, where $l_E$ is the field transverse space in the polarization direction and $F_0$ its characteristic amplitude in Schwinger field units, both safely met in our cases of study, this can be done integrating the local production rate (Eq.~\ref{Scw_rate}) in the full simulated spacetime volume \cite{Cohen2008} . 

In Fig.~\ref{Fig7} we show the resulting numbers for the field configurations of a single focused Gaussian pulse, two counter-propagating Gaussian pulses (all supplemented with Gaussian time envelopes as in previous section), and for the optimal harmonic beam. All beams are focused close to their diffraction limit ($\varepsilon=1/\pi$) and only their energy (amplitude) is varied. In all cases the number of pairs grows exponentially with the intensity, and thus increases very fast past the one pair creation threshold. The corresponding threshold intensity however varies widely with the field configuration. For two counter-propagating Gaussian pulses, pair creation becomes significant at intensities of $I_0\simeq 2.4\times 10^{27} \mathrm{\ W.cm^{-2}} \ (E=7.21\times10^{-2}\ E_S)$, two orders of magnitude below the Schwinger intensity, in agreement with \cite{FedotovNarozhni2014_pairs_rev,NarozhniBulanov2004_ctrGaus_first,Bulanov2006_EPpairs,BulanovNarozhny2006_1Gaus_2ctrGaus_tables}. As noticed in the previous studies, this is possible in spite of the local production rate suppression, due to the macroscopic size of the interaction volume compared to the electron Compton scale. On the other hand, in a single beam configuration we find that intensities of $I_0\simeq 7.5\times 10^{28} \mathrm{\ W\,cm^{-2}} \ (E=7.03\times10^{-1}\ E_S)$ are required for a similar interaction volume, in agreement with \cite{Fedotov2009_Stronglyfoc,BulanovNarozhny2006_1Gaus_2ctrGaus_tables}, as a consequence of the same invariants suppression discussed previously for photon-photon scattering. In this regard the smaller interaction volume of the harmonic beam explains its even higher threshold of $I_0\simeq 6.4\times 10^{29} \mathrm{\ W\,cm^{-2}} \ (E=1.18\ E_S)$. For that matter, we note this value implies that exceeding the critical field value on macroscopic space extents is in principle possible within the CHF scheme. This is in contrast with other beam configurations tailored for enhancing vacuum pair production and subsequent QED cascades, such as counter propagation geometries, where efficient light to matter conversion once at least one pair is created effectively limits the intensity attainable with such fields \cite{Bell2008,Fedotov2010,Bulanov2010,Nerush2011,Grismayer2016}.

Assuming fixed focusing at diffraction limit in all cases, we can then convert these intensity thresholds into requirements on the power of the driving beams, which offers a basic assessment of experimental feasibility. In order to avoid unnecessary modelling specifics we define power as $P_0\equiv W/\tau$, where the total beam energy $W$ is our simulation input and $\tau=25$ fs, matching the duration duration of the considered Gaussian pulses, and the approximate harmonics train duration. The power values determined along these lines correspond either to the sum of the powers of the individual beams in the case of counterpropagating pulses, to the total beam power in the single pulse case, and to the hypothetical power of the laser pulse before PM interaction in the case of the harmonic beam, for a realistic reflection coefficient of $70\ \%$ \cite{Vincenti_PMopt}. We assume in particular that the harmonic spectrum remains independent of the total field energy. The latter assumption is an idealization of course, however it does not strictly rely on the invariance of laser-plasma interactions over a wide intensity range. Indeed one could in principle simply increase the laser waist on plasma mirror so as to maintain a quasi-constant incident intensity; for instance, a 200 PW laser with incident waist $w_p=25\lambda_0$ reaches $a_0\simeq115$, only about $50\ \%$ more than a 3 PW laser with $w_p=4.5\lambda_0$ ($a_0\simeq75$) as used in the PIC simulation from which the spectrum was extracted. From this perspective, CHF clearly emerges as the most favorable configuration, as Schwinger pair production first occurs for laser powers between 160 PW and 200 PW ($W_{las}\lesssim 5$ kJ at 25 fs), barely beyond reach of the most powerful facilities planned to date \cite{PetaExaWW2019}, and allowing a single-beam setup. According to our computation, achieving the same without harmonics conversion would require close to 2.6 EW cumulative power with two counter-propagating pulses ($W_{las}=65$ kJ), a value that could potentially be reduced to 1.1 EW ($W_{las}=27$ kJ) \emph{via} the coherent combination of more individual beams in the so-called dipole wave setup \cite{Dipole_pairs}. The latter requirements remain about one order of magnitude beyond the currently envisioned facilities. Using a single optical pulse does not stand as a viable option in this respect, with a required power of 200 EW ($W_{las}=6.4$ MJ).

\section{Vacuum signatures from a secondary beam assisted CHF field}
\label{secIV}

\begin{figure}[t!]
\begin{center}
\includegraphics[width=1.0\columnwidth]{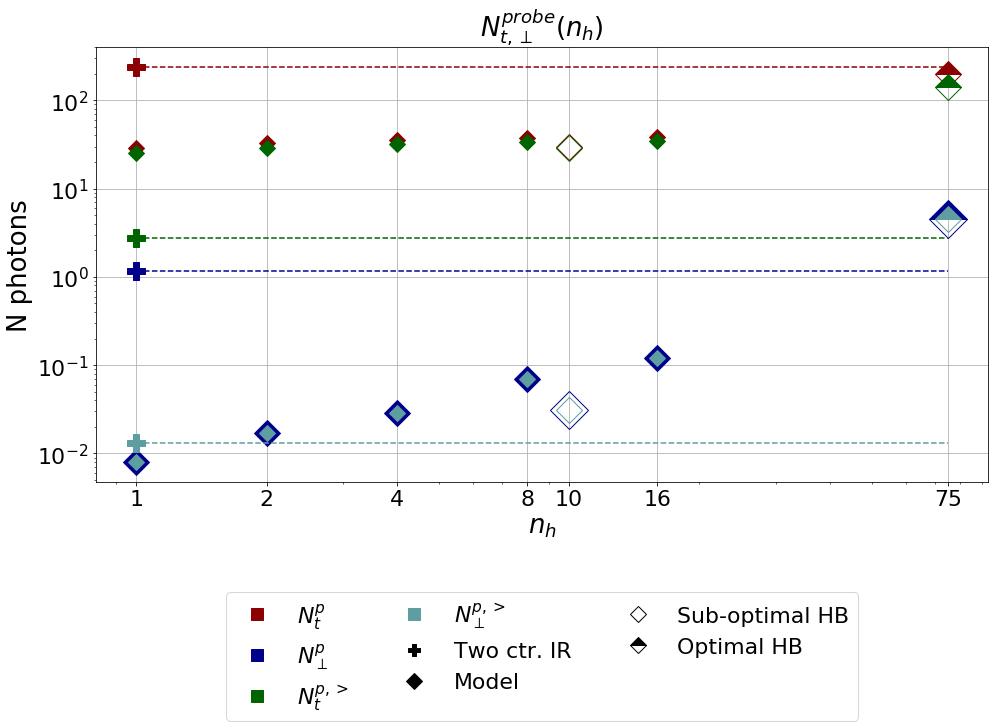}
    \vspace{-20pt}
     \caption{\textbf{Number of photons from a loosely focused beam scattered off a counter-propagating beam made of $n_h$ harmonics of its fundamental frequency, or PM-generated beams with effective number of harmonics $n_h$, focused at diffraction limit.} Except from \textbf{(2 ctr. Gaus.)}, all shown number correspond to photons scattered in the half-space towards which the probe beam is directed, as denoted by a superscript $p$. The model beams are constructed as in sec. III. The PM-generated harmonic beams are linearly polarized at an angle $\psi_t=\pi/2$ compared to the probe beam main polarization direction.
     }
     \label{Fig8}
\end{center}
\end{figure}


As demonstrated in \cite{FK_AllOpt2019,Asm_numerics} and consistently with analytical estimates, the numbers of signal photons emitted from a Gaussian beam colliding with a plane wave at an angle with the same linear polarization are maximized for a counter-propagating geometry, and vary as $(1-cos(\theta_{collision}))^4$ around this optimal collision angle \cite{FK2021}. Besides, as explicited in sec.~\ref{secIII}, a counter-propagating geometry gives rise to nonvanishing invariants, which strongly enhances photon emission over a single focused beam situation. We therefore investigate here the coupling of the focused PM-generated beam, thought of as a high-intensity ``pump'' for the quantum vacuum, with a loosely focused counter-propragating single frequency beam, thought of as a well-controlled ``probe'' to be scattered off the intense, strongly localized harmonic beam focal spot. This can \emph{a priori} help to increase the \emph{discernible} photon number in two ways, namely \emph{i)} by emitting some signal photons, which can be interpreted as scattered probe photons, away from the plasma mirror specular direction likely polluted by background plasma radiations, and \emph{ii)} by increasing the scattering angles of the probe photons, due to the high transverse momenta of the harmonic photons. In a semi-classical picture the latter effect can be explained by diffraction of the probe beam off the intense pump field polarizing the quantum vacuum in a sub-wavelength region. For these reasons all photon numbers given in the following correspond to "scattered probe photons", \emph{i.e.} we completely discard photons scattered in the half ($\mathbf{k}-$)space around the harmonic beam propagation direction from the outset. We consider two different options for the probe, namely either (IR)
a loosely focused Gaussian pulse with $\lambda_0 = 800$ nm, $\tau = 25$ fs, $I_0=2.5\times10^{21}$ W.cm$^{-2}$ and $w_0 = 9\lambda_0$, or (Gr) an even less focused green laser with $\lambda_g = 527$ nm, $\tau = 100$ fs, $I_0 = 3.5\times10^{19}$ W.cm$^{-2}$ and $w_0 = 14 \lambda_g$. The first one can be thought of as a split part of a multi-Petawatt laser, whose second part would be driving the harmonic generation. The second option would typically involve an auxiliary 100 TW green laser close to the one suggested in \cite{DiPiazza2010_matterless}. In both cases, two polarizations of the probe beam are envisioned, both linear, with an angle $\psi=\sphericalangle(\mathbf{e}_{pr},\mathbf{e}_{h})$ with the harmonics polarization direction of either $\psi_t = \pi/2$ or $\psi_\perp = \pi/4$, aimed at maximizing the total or crossed-polarized photon numbers respectively \cite{FK2019_xray_ph_scatt} (in this context ``crossed-polarized'' means polarized in the direction perpendicular to the main polarization direction of the probe beam).

In order to clarify to what extent can CHF improve over the single frequency beam signal in this configuration, we first computed the probe photon numbers scaling with the number $n_h$ of harmonics, using the same model harmonic beam as in sec.~\ref{secIII}, colliding with the IR secondary beam. 

These values show a slow increase in total numbers of photons, and a steeper increase in crossed-polarized photons; in the optimal harmonic beam case, $N^p_t$ reaches levels comparable to the one of two counter-propagating beams at diffraction limit while $N^p_\perp$ exceeds those. Most importantly though, the number of discernible photons emitted outside of a $\theta = \arcsin(2\varepsilon_b)$ cone around the probe beam direction is almost equal to the total number of probe scattered photons. This is due mostly to the small probe beam opening angle, as the effect persists even in the single frequency case ($n_h=1$ in Figure \ref{Fig8}). The results for the PM-generated beams with varying probe beam parameters are displayed in Table 1. It confirms the very high number of probe photons scattered at large angles compared to the probe beam divergence, and shows that varying the probe beam polarization effectively allows to optimize total or crossed-polarized photon numbers in most cases. 

\begin{table}[t]
\begin{center}
    \begin{tabular}{|c|c|c|c|c|c|c|}
        \hline
         HB spectrum & probe & $\sphericalangle(\mathbf{e}_{pr},\mathbf{e}_{h})$ & $N^p_t$ & $N^p_\perp$ & $N^{p,>}_t$ & $N^{p,>}_\perp$ \\
        \hhline{|=|=|=|=|=|=|=|}
       sub-optimal  & IR    &  $\psi_t$      &  29.6  &   3.11$\times10^{-1}$  &    29.1 &   3.10$\times10^{-2}$  \\
       \cline{3-7}
                &       &  $\psi_\perp$  &  19.6  &   1.38  &  19.3  &   1.35  \\
       \cline{2-7}
                & Gr    &  $\psi_t$      &  1.43  &   7.72$\times10^{-3}$  &    1.40 &   7.72$\times10^{-3}$   \\
       \cline{3-7}
                &       &  $\psi_\perp$  &  9.58$\times10^{-1}$  &   7.49$\times10^{-2}$  &  9.37$\times10^{-1}$  &   7.34$\times10^{-2}$  \\
       \hhline{|=|=|=|=|=|=|=|}
       optimal  & IR    &  $\psi_t$      &  199  &  4.58  &  142 $\vert$ $9.49_{2\omega_0}$  &  4.58 $\vert$ $2.61_{2\omega_0}$  \\
       \cline{3-7}
                &       &  $\psi_\perp$  &  132  &  10.5  &  94.2 $\vert$ $6.31_{2\omega_0}$ & 7.86 $\vert$ $2.37_{2\omega_0}$   \\
       \cline{2-7}
                & Gr    &  $\psi_t$      &  24.7 &  4.16  &  24.5 $\vert 1.07_{\omega_g+\omega_0}$  &  4.16   $\vert$ $0.19_{\omega_g+\omega_0}$ \\
       \cline{3-7}
                &       &  $\psi_\perp$  &  16.9  &   2.25  & 16.8 $\vert 0.70_{\omega_g+\omega_0}$   &    2.22 $\vert 0.14_{\omega_g+\omega_0}$\\
        \hline
    \end{tabular}
    \caption{\textbf{Number of probe beam photons scattered in the collision with a PM-generated harmonic beam, for different harmonics spectrum, probe beam and polarization directions.} The subscripts stand for numbers of probe photons emitted in the range $\Delta\omega = \omega_0/2$ around the indicated frequency.}
\end{center}
\vspace{-20pt}
\end{table}

The final numbers of discernible signal photons then appear significantly higher than in all the other configurations studied here. More specifically, resorting to CHF allows at most for a gain of almost two orders of magnitude in discernible photon numbers compared to directly using a split infrared pulse as both the pump and the probe for vacuum polarization effects (comparing optimal HN and two ctr. Gaus. in Fig~\ref{Fig8}). Admittedly this achievement still has to be weighed against the additional experimental requirements of PM harmonics generation, however, we can identify two aspects in which the CHF scenario stands apart from simpler configurations. 

First we note that the numbers of crossed-polarized photons benefits from even greater enhancements than the polarization insensitive photon numbers, reaching values higher than in any other configuration studied here by one to almost three orders of magnitude, respectively for the total and discernible populations (Figure~\ref{Fig8}). 

Second and most importantly, a significant fraction of the probe photons undergo \emph{inelastic} scattering with energy gains of $+\omega_0$, which corroborates the findings of \cite{BoostCHF2019}. In the case of the IR probe beam and optimal spectrum, this fraction can go up to $6.7\ \%$ in total, and $1.8\ \%$ of all the emitted photons are both crossed-polarized and have a $2\omega_0$ frequency, which amounts to more than 2 photons per shot. With the Gr probe beam and $\psi_t$ polarization angle, more than one photon per shot is emitted in the $\omega_g+\omega_0 \simeq 2.5\omega_0$ range, and one every five shots adding the crossed-polarization constraint. Such photon signals are in principle detectable in experiments \cite{doyle2021}. For reference, in the case of two counter-propagating pulses the proportion of inelastically scattered photons (at $3\omega_0$) is of the order of $0.01\ \%$, and the model harmonic beam in collision with the IR probe yields about $0.3\ \%$ (at $2\omega_0$). This indicates that inelastic scattering in the probe beam direction is inherent to the interaction with a multi-frequency beam. 

This property can potentially prove decisive for obtaining a signal discernible from the background of the driving laser photons beyond reasonable doubt. Indeed, the discernibility criterion used in our work as well as in many others \cite{FK_SVE2015,FK_AllOpt2018,FK_Multicolor} typically relies on the idealization of the photon distribution of the background beams as Gaussian, hence falling to negligible levels at a finite separation from their propagation direction. However, as there are about 20 orders of magnitude more background than signal photons, this assumption is extremely vulnerable to any laser imperfections or insufficient shielding against parasitic radiation \cite{doyle2021}, even at crossed-polarization. On the opposite, a large frequency separation could allow establishing the quantum origin of a photon on a much firmer ground. 

\section{Conclusion}
\label{Concl}

Observing quantum vacuum processes with electromagnetic fields typically requires approaching the Schwinger field $F_S = 1.32\times10^{18} {\rm\ V\,m^{-1}} = 4.4\times10^9 {\rm\ T}$, several orders of magnitude above current technological capabilities, and therefore pleads for any scheme maximizing the attainable field intensity. On the other hand, Lorentz invariance of the vacuum state implies that all the relevant expectation values depend on the Lorentz invariants of the field rather than on its amplitude, which generically explains their strong dependence upon the interaction geometry, and thus raises the question of the optimal configuration beyond achieved intensity. In this work, we have shown that optimal focusing of the harmonic beam produced in the reflection of a single Petawatt-class laser off a plasma mirror allows to generate as many vacuum photons as two perfectly counter-propagating Petawatt infrared pulses focused at diffraction limit and to significantly reduce the laser powers required to observe Schwinger pair creation.

The physical relevance of this result nevertheless crucially depends on the degree of control of the laser-plasma interactions achievable in an experiment. Besides the requirements of optimal harmonics generation and focusing \cite{quere_vincenti_2021}, the level and nature of background plasma mirror emissions prevails as to which observations will be accessible. If the total radiated field after plasma mirror interaction can be consistently assimilated to the reflected harmonic beam, with sharp enough angular photon distribution, propagating in a vacuum, as we modelled it in this work, then it would be possible to detect vacuum photon scattering with the coherent harmonics focusing technique only. The coupling to a well controlled auxiliary beam significantly alleviates these constraints, only requiring low background emissions in the "probe" beam direction, which could be set close to counter-propagation with the harmonic beam. Furthermore, due to the high levels of inelastically scattered probe photons specific to this case, assessment of the quantum origin of the signal could then be made on a spectral basis, hence with potentially much higher confidence than solely on the basis of an angular discernibility criterion. As associated energy gains are of the order of $+\omega_0$, use of a $\omega_p\neq\omega_0$ frequency probe can result in even cleaner quantum vacuum signatures in the $\omega_p + \omega_0$ frequency range. Indeed, provided the probability of presence of residual charged particles in the overlap region of the beams can be made small enough such frequencies could not be generated by laser-matter interaction. 

Irrespective of the chamber residual gas, whose density can be made small enough to empty the harmonics focal spot with high probability, the near-specular emission of relativistic electron beams directly from the plasma mirror surface is an experimentally established fact \cite{Thevenet2016,Chopineau2019}, so that determining whether the above conditions can be met calls for a detailed study of these ejected electrons beyond the scope of this work. If these electrons are expelled from the harmonics field early enough, photon-photon scattering may still be observed.
If on the opposite they radiate enough to preclude observation of photon-photon scattering, but do not trigger QED cascades, quantum vacuum processes could still be sought for at higher intensity in the form of Schwinger pair creation.
If finally the electron beam dynamics results in prolific pair creation even before the Schwinger process can occur, detection of quantum vacuum effects from PM-generated beams will likely require more complex setups, typically involving disentanglement of the generation and focusing steps; in turn, direct interaction of the curved PM harmonics beam with matter, either in the form of beams or of a secondary target \cite{Fedeli2021}, would then open the way to yet unobserved regimes of plasma dynamics in strong-fields.   

\clearpage

\appendix

\section{Computational resources}
\label{App_comp}

The general lack of symmetry of the source terms in Eq.~\ref{Tf_PM} makes the algorithm described in sec. 2 potentially very expensive from the computational point of view. Indeed, working with 3-dimensional numerical arrays we can infer memory needs of $\mathcal{M}\propto N_xN_yN_z$, while the time complexity stemming from the on-shell Fourier transform is $\mathcal{T}\propto N_tN_xN_yN_z \log(N_xN_yN_z)$. On the other hand, the resolved scales in each space-time dimension need to range from the field support characteristic length $1/L$, in order to simulate the whole relevant interaction volume, to $2\times \nu_{max}$, in order to fulfil the Shannon criterion associated to the maximal field frequency, as scattering events with significant energy exchanges are suppressed in comparison to the elastic channels \cite{FK_SVE2015,FK_AllOpt2018}. For a focused monochromatic Gaussian pulse with waist at focus $w_0 \sim\lambda_0$ and duration $\tau \sim 10 \lambda_0$, choosing a minimal (maximal) propagation time such that the ``front'' (``rear'') of the pulse is at about three Rayleigh lengths from focus implies that fitting the whole pulse in the simulation space at all times requires a total longitudinal length of about $2 \times (\tau + 3z_R) \simeq 40\lambda_0$, and a transverse length close to $20\lambda_0$ as a consequence of the large divergence of a tightly focused beam. If we furthermore assume the need to work with about ten 3-dimensional double precision (8 Bytes) arrays at any given moment (\emph{e.g.} from field components, Lorentz invariants, etc.), we are then led to,
\begin{align}
    \mathcal{M}_{Gaus} &\sim 10\times(2\times \nu_0 \times 40\lambda_0)\times(2\times \nu_0 \times 20\lambda_0)^2\times 8 \text{ Bytes}\\
    & \sim 10 \text{ MB}
\end{align}
This is a rather modest memory requirement, making such configurations computable on a laptop. If we now consider a beam comprising harmonics up to order $n_h \sim 100$, the maximum frequency is of course $\nu_{max}\sim 100 \nu_0$, while the field falling length is still bound to the global duration of the beam $\tau\sim 10\lambda_0$, including in the transverse dimensions again due to the large divergence of a very focused beam, which implies,
\begin{align}
    \mathcal{M}_{h} &\sim 10\times(2\times 100\nu_0 \times 20\lambda_0)\times(2\times 100\nu_0 \times 10\lambda_0)^2\times 8 \text{ Bytes}\\
    & \sim 100 \text{ GB}
\end{align}
Combined with the time constraint, this amounts to a computational volume of about $2.5\times10^4$ CPUhours/run, and thus marks the need for a large scale computing infrastructure.

\section{Harmonics spectra from PIC simulations}
\label{App_harmonics}

\begin{figure}[t!]
\begin{center}
\includegraphics[width=1.0\columnwidth]{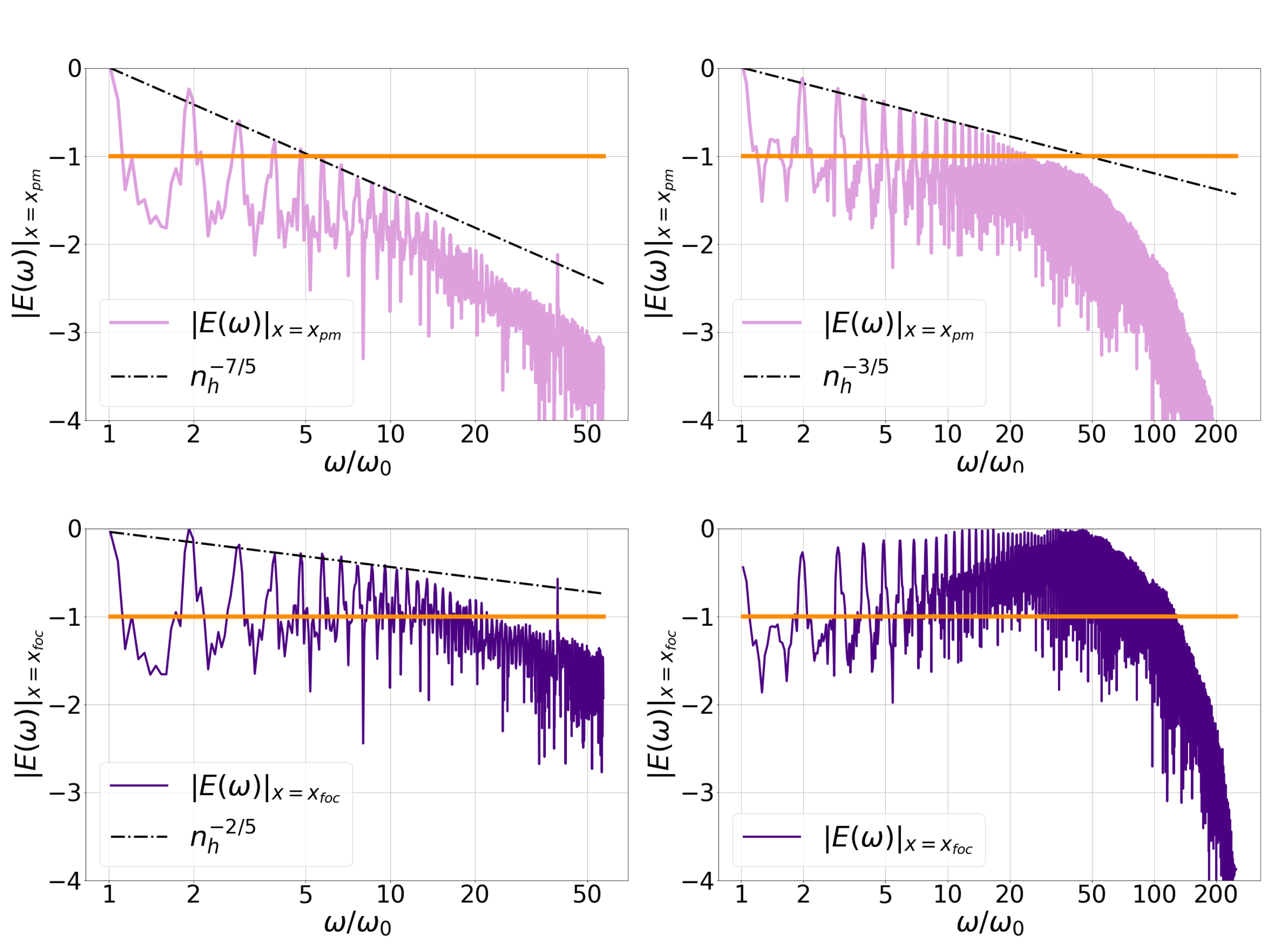}
     \caption{\textbf{Time spectra of the main component of the electric field on-axis extracted from Particle-In-Cell plasma mirror simulations (logscale).} Original spectra of the sub-optimal \textbf{(left)} and optimal \textbf{(right)} harmonic beams are recorded in the plasma mirror plane \textbf{(top)}, while the resulting spectra at focus \textbf{(bottom)} are shown assuming perfect focusing. The $10^{-1}$ level is shown (orange line) so as to provide a proxy for the cutoff harmonic order.
     }
     \label{Fig_spectra}
\end{center}
\end{figure}

Under suitable interaction conditions, an overdense plasma irradiated by an ultra-intense laser pulse effectively acts as a mirror of optical quality, affecting essentially only the time spectrum of the field from the Doppler effect caused by its relativistic oscillations, without inducing major optical aberrations.

In these conditions, it is legitimate to approximate the reflected field by a superposition of monochromatic components each having a Gaussian spatial profile. Then the field is entirely determined by the data of $ w_p(\omega)$, $R_p(\omega)$ and $A(\omega)$, respectively the (frequency-dependent) beam waist, radius of curvature and time spectrum of the main component of the electric or magnetic field, all taken in the PM plane. In this work we assumed that $w_p$ and $R_p$ were the same for all frequencies, implying they only encode the opening angle of the harmonic beams (the diffraction parameter then being equal to $\varepsilon = w_p/R_p$ if we assume $\omega w_p\gg1 $ for all frequencies $ \omega$). The non trivial information about the PM harmonics generation is then entirely contained in the frequency spectrum $A(\omega)$, including its maximal attainable amplitude for a given field energy. This critical data was provided by Particle-In-Cell simulations using the Warp+PICSAR framework \cite{WarpPIC2012,WarpPIC2013,WarpPIC2016,WarpPIC2018}.

As specified in the main text, we used two spectra in this study, corresponding either to sub-optimal or optimized parameters of the laser-plasma interactions generating the harmonics field. The modulus of these numerical spectra is shown in Figure~\ref{Fig_spectra}, in the PM plane and at focus. The PM plane data corresponds to the output of the PIC simulation. In order to extract the propagative component of the field from these simulations, field values are actually recorded in a plane away from the PM surface, and then backpropagated on PM surface by vacuum Maxwell's equations. Spectra at PM focus are calculated from the PM plane spectra assuming perfect focusing, \emph{i.e.} $E(\omega)\vert_{x=x_{foc}} \propto \omega \times E(\omega)\vert_{x=x_{PM}}$. It then appears that slowly decaying spectra can be obtained, with roll-off parameters (see sec.~\ref{secIII}) of $s=7/5$ and $s=3/5$, with harmonics numbers of the order of a few tens in PM plane, resulting in quasi-constant harmonics amplitude at PM focus up to an harmonic order of more than 50 in the optimal case. 


\section*{Acknowledgments}
The authors would like to thank Fabien Quéré for fruitful discussions in the initiation and development of this work. This research used resources of the Oak Ridge Leadership Computing Facility at the Oak Ridge National Laboratory, which is supported by the Office of Science of the U.S. Department of Energy under Contract No.DE-AC05-00OR22725. We acknowledge the support of the US DOE Exascale Computing Project and of the Director, Office of Science, Office of High Energy Physics, of the U.S. Department of Energy under Contract No. DEAC02-05CH11231. FK has been funded by the Deutsche Forschungsgemeinschaft (DFG) under Grant No. 416607684 within the Research Unit FOR2783/1.


\end{document}